\documentclass[12pt]{article}
\usepackage{amsxtra,amssymb,amsthm,amsmath,latexsym}

\theoremstyle{plain}

\def\R{{\mathbb R}}

\def\C{{\mathbb C}}

\def\oH{{\overset{\circ}{H}}}
\def\oH1{{\overset{\circ}{H}\kern-.02in{}^1}}

\def\bee{\begin{equation*}}
\def\eee{\end{equation*}}
\def\be{\begin{equation}}
\def\ee{\end{equation}}

\begin{document}

\title{On the denseness of the set of scattering amplitudes
}

\author{Alexander G. Ramm\\
 Department  of Mathematics, Kansas State University, \\
 Manhattan, KS 66506, USA\\
ramm@math.ksu.edu\\
http://www.math.ksu.edu/\,$\widetilde{\ }$\,ramm
}

\date{}
\maketitle\thispagestyle{empty}

\begin{abstract}

It is proved that the set of scattering amplitudes $\{A(\beta, \alpha, k)\}_{\forall \alpha \in S^2}$, known for all $\beta\in S^2$, where $S^2$ is the unit sphere in $\mathbb{R}^3$,  $k>0$ is fixed, $k^2$ is not a Dirichlet eigenvalue of the Laplacian in $D$,
is dense in $L^2(S^2)$.  Here $A(\beta, \alpha, k)$ is the scattering amplitude corresponding to an obstacle $D$, where $D\subset \mathbb{R}^3$ is a bounded domain with a boundary $S$.
 The boundary condition
on $S$ is the Dirichlet condition.
\end{abstract}

\section{Introduction}\label{S:1}
 Let $D\subset \mathbb{R}^3$ be a bounded domain with a connected $C^2-$smooth boundary $S$,
 $D':=\mathbb{R}^3\setminus D$ be the unbounded exterior domain and $S^2$ be the unit sphere in $\mathbb{R}^3$.

Consider the scattering problem:
\be\label{e1}
(\nabla^2+k^2)u=0 \quad in\quad D', \qquad u|_{S}=0, \qquad u=e^{ik\alpha \cdot x}+v,
\ee
where $k>0$ is a constant, $\alpha\in S^2$ is a unit vector in the direction of the propagation of the incident plane wave $e^{ik\alpha \cdot x}$,  $N$ is the unit normal to $S$ pointing out of $D$,
$u_N$ is the normal derivative of $u$,   and
 the scattered field $v$ satisfies the radiation condition
\be\label{e2} v_r-ikv=o\Big(\frac 1 r\Big), \quad r:=|x|\to \infty.
\ee
The scattering amplitude $A(\beta, \alpha, k)$ is defined by the following formula:
\be\label{e3}
v=A(\beta, \alpha,k)\frac{e^{ikr}}{r} + o\Big(\frac 1 r\Big), \quad r:=|x|\to \infty, \quad \frac {x}{r}=\beta,
\ee
where $\alpha, \beta \in S^2$, $\beta$ is the direction of the scattered wave, $\alpha$ is the direction of the incident wave.
For a bounded domain $ o(\frac 1 r)=O(\frac 1 {r^2})$ in formula \eqref{e3}.
The function $A(\beta, \alpha, k)$ is the scattering amplitude.
It is known (see \cite{R190}, p.25) that the solution to the scattering problem \eqref{e1}-\eqref{e3} does exist and is unique.

It is of basic interest to many physical problems to know whether the set $\{A(\beta, \alpha, k)\}_{\forall \alpha \in S^2}$ is dense in $L^2(S^2)$. For example, the book \cite{K} discusses the question of the possibility to approximate a given diagram (that is, a given function in $L^2(S^2)$) by the diagram generated by a distribution of current on a surface antenna. Examples show that this is not always possible. 

For example, let the scattering amplitude be of the form (see lemma 3 below):
$$A( \beta, \alpha, k)=-\frac 1 {4\pi} \int_S e^{-ik\beta \cdot s}h(s, \alpha, k)ds, \quad h:=u_N,$$
where $u_N$ is the normal derivative of $u$ on $S$, and $u=0$ on $S$. 
This is the scattering amplitude for acoustic scattering of a plane wave by an acoustically soft body.
The following result, essential for our paper, is formulated in Lemma 2 below: 
 for a fixed $k>0$,  the set $\{h\}|_{\forall \alpha \in S^2}$ is dense in $L^2(S)$.

Then the question is: can one
approximate  an arbitrary function $f(\beta)\in L^2(S^2)$
 in $L^2(S^2)$ norm by the scattering amplitude $-\frac 1 {4\pi} \int_S e^{-ik\beta \cdot s}h(s, \alpha, k)ds$ 
 with an error not exceeding $\epsilon >0$, where $\epsilon$ is an arbitrary small given number and $k>0$ is fixed?

The purpose of this paper is to prove Theorem 1 which answers the above question by giving a sufficient condition for
the possibility of such an approximation. 

{\bf Theorem 1.}  {\em Assume that $k>0$ is fixed and $k^2$ is not a Dirichlet eigenvalue of the Laplacian in $D$. Then the set
$\{A(\beta, \alpha, k)\}_{\forall \alpha \in S^2}$ is dense in $L^2(S^2)$. }
\vspace{3mm}

In Section 2 three lemmas are formulated and  Theorem 1 is proved.

\section{ Proof of Theorem 1}\label{S:2}

{\bf Lemma 1.} (\cite{R190}, p. 46) {\em One has:
\be\label{e4}
G(x,y,k)= g(|y|)u(x,\alpha,k) + O\Big(\frac 1 {|y|^2}\Big), \qquad |y|\to \infty, \quad \frac y{|y|}=-\alpha.
\ee
}
Here $g(|y|):=\frac{e^{ik|y|}}{4\pi |y|}$,  $u(x,\alpha,k)$ is the scattering solution, i.e.,
the solution to problem \eqref{e1}-\eqref{e3}. Lemma 1 gives a new definition of the scattering solution  $u(x,\alpha,k)$.

{\bf Lemma 2.} (\cite{R470}, p. 242) {\em Let $k>0$ be fixed. The set $\{u_N(s,\alpha,k)\}|_{\forall \alpha \in S^2}$ is dense in $L^2(S)$. }
\vspace{3mm}

{\bf Remark 1.} The conclusion of Lemma 2 is valid without the assumption that
$k^2$  is not a Dirichlet eigenvalue of the Laplacian in $D$. Lemmas 1 and 2 are proved by the author
and they allow to give a short proof of Theorem 1.

\vspace{3mm}

{\bf Lemma 3.} {\em One has
\be\label{e5} -4\pi A(\beta, \alpha, k)=\int_S e^{-ik\beta\cdot s} u_N(s,\alpha,k)ds.
\ee
}

\vspace{3mm}

{\bf Proof of Theorem 1.} Let us assume that a function $f(\beta)$ exists, such that
\be\label{e6}
\int_{S^2} A(\beta, \alpha, k)f(\beta)d\beta=0 \qquad \forall \alpha\in S^2.
\ee
Using Lemma 3 one gets  (remember that $k$ is a constant):
\be\label{e7}
\int_S ds u_N(s,\alpha, k)) \int_{S^2}d\beta e^{-ik\beta \cdot s} f(\beta)d\beta=0 \qquad \forall \alpha\in S^2,
\ee
so, by Lemma 3, one concludes that
\be\label{e8}
\int_{S^2} e^{-ik\beta \cdot s} f(\beta)d\beta=0   \qquad \forall s\in S.
\ee
Define
\be\label{e9}
w(x):=\int_{S^2}  e^{-ik\beta \cdot x} f(\beta)d\beta.
\ee
This is an entire function of $x$, and
\be\label{e10}
(\nabla^2+k^2)w=0  \quad in \quad D, \qquad w|_S=0.
\ee
If $k^2$ is not a Dirichlet eigenvalue of the Laplacian in $D$, then \eqref{e10} implies $w=0$ in $D$,
and, by analyticity of $w$, it follows that $w=0$ in $\R^3$.
Therefore $f(\beta)=0$. Indeed, it follows from  \eqref{e9} that Fourier transform of the distribution
$\delta(\lambda -k)(\lambda)^{-2}f(\beta)$ is zero. The Fourier transform variable $\xi=\lambda \beta$,
where $\lambda:=|\xi|$, $\beta:=\xi/\lambda$.
Thus, by the injectivity of the Fourier transform, this distribution is zero, so $f(\beta)=0$.
Theorem 1 is proved. \hfill$\Box$

{\bf Remark 2.} If $S$ is a sphere of radius $a$ centered at the origin and $j_{\ell}(ka)=0$, where
$\ell\ge 0$ is an integer, and $j_l(r)$ is the spherical Bessel function (see, for example, \cite{R470}, p.262), then $k^2$ is a Dirichlet eigenvalue of the Laplacian in the ball of radius $a$ centered at the origin. In this case the set
 $\{A(\beta, \alpha, k)\}_{\forall \alpha \in S^2}$ is orthogonal to $Y_{\ell}$, and, therefore, is not
 dense in $L^2(S^2)$.

A simple calculation proves the above conclusion. It is well known that
\be\label{e11}
e^{-ik\beta \cdot s}=\sum_{\ell=0}^\infty 4\pi (-i)^{\ell} j_{\ell}(k|s|) \overline{Y_{\ell}(\beta)}Y_{\ell}(s^0), \quad s^0:=s/|s|,
\ee
where $Y_{\ell}$ are the spherical harmonics (see \cite{R470}, p.261), and the summation with respect to $\ell$ means
 the summation with respect to $\ell$ and $m$, $-\ell\le m \le \ell$, see \cite{R470}, p. 261. From \eqref{e8} and \eqref{e11} it follows
for the case when $S$ is a sphere of radius $a$ that
\be\label{e12}
\sum_{\ell=0}^\infty 4\pi (-i)^{\ell} j_{\ell}(k|a|)Y_{\ell}(s^0)f_\ell=0,
\ee
where $f_\ell:=\int_{S^2} f(\beta)\overline{Y_{\ell}(\beta)}d\beta$.
It follows from \eqref{e12} that $f_\ell=0$ for all $\ell$ except the $\ell_0$ for which $j_{\ell_0}(ka)=0$.
For this $\ell_0$ the coefficient $f_{\ell_0}$ can be arbitrary. Therefore the set
$\{A(\beta):=A(\beta, \alpha, k)\}_{\forall \alpha \in S^2}$ is orthogonal to $Y_{\ell_0}(\beta)$.

In \cite{K} the following question is discussed:

Is it true that one can approximate an arbitrary
function in $L^2(S^2)$ by a far-field diagram generated by a distribution of sources (currents)
on a given surface $S$?

The author of \cite{K} shows that in two-dimensional problems it is not always possible to approximate an arbitrary
function, belonging to $L^2(S^2)$ by a far field. From Theorem 1 it follows that this conclusion is true only if
$k^2$ is a Dirichlet eigenvalue of the Laplacian in $D$, where $S$ is the boundary of $D$. In \cite{K} there is also
a statement that there are "very many" such surfaces, but this statement is not proved. The author thinks that this statement
is true.

It is an interesting open
problem to find out if there are surfaces $S$ which are not spheres of radius $a$, centered at the origin,
on which the entire function \eqref{e9} vanishes. This question is similar to the symmetry problem
related to Pompeiu problem, \cite{R666}

 A possible approach to this problem can be outlined. The function $w$, defined in \eqref{e9}, is an entire function of $x$.
It solves equation \eqref{e10} in $\R^3$, and, in particular, in $D$, and $w=0$ on $S$. The set on which $w=0$ is an
analytic set in $\C^3$ which intersection with $\R^3$ is the surface $S$.
A small perturbation of $f(\beta)$ in \eqref{e9} leads to a small perturbation of the above analytic set, and, therefore,
to a small perturbation of $S$. So, varying $f$ one can vary $S$.


\begin{thebibliography}{1000} 

\bibitem{K} B.Z. Katsenelenbaum, {\em Problems of approximation of electromagnetic field},
 Nauka, Moscow, 1996.  (in Russian)


\bibitem{R190} A.G.Ramm,  {\em Scattering by obstacles}, D.Reidel, Dordrecht, 1986.

\bibitem{R470} A.G.Ramm, {\em Inverse problems}, Springer, New York, 2005.

\bibitem{R666} A.G.Ramm,  Solution to the Pompeiu problem and the related symmetry problem,

Appl. Math. Lett., 63, (2017), 28-33.





\end{thebibliography}
\end{document}